\documentclass[11pt]{article}
\usepackage{graphicx,epsf,tikz,wrapfig,eufrak}
\usepackage{verbatim,amsmath,eufrak}
\usepackage{cite} 

\renewcommand{\arraystretch}{1.2}
\makeatletter
\newdimen\normalarrayskip              
\newdimen\minarrayskip                 
\normalarrayskip\baselineskip
\minarrayskip\jot
\newif\ifold             \oldtrue            \def\new{\oldfalse}
\def\arraymode{\ifold\relax\else\displaystyle\fi} 
\def\eqnumphantom{\phantom{(\theequation)}}     
\def\@arrayskip{\ifold\baselineskip\z@\lineskip\z@
     \else
     \baselineskip\minarrayskip\lineskip2\minarrayskip\fi}
\def\@arrayclassz{\ifcase \@lastchclass \@acolampacol \or
\@ampacol \or \or \or \@addamp \or
   \@acolampacol \or \@firstampfalse \@acol \fi
\edef\@preamble{\@preamble
  \ifcase \@chnum
     \hfil$\relax\arraymode\@sharp$\hfil
     \or $\relax\arraymode\@sharp$\hfil
     \or \hfil$\relax\arraymode\@sharp$\fi}}
\def\@array[#1]#2{\setbox\@arstrutbox=\hbox{\vrule
     height\arraystretch \ht\strutbox
     depth\arraystretch \dp\strutbox
     width\z@}\@mkpream{#2}\edef\@preamble{\halign
\noexpand\@halignto
\bgroup \tabskip\z@ \@arstrut \@preamble \tabskip\z@ \cr}%
\let\@startpbox\@@startpbox \let\@endpbox\@@endpbox
  \if #1t\vtop \else \if#1b\vbox \else \vcenter \fi\fi
  \bgroup \let\par\relax
  \let\@sharp##\let\protect\relax
  \@arrayskip\@preamble}
\def\eqnarray{\stepcounter{equation}%
              \let\@currentlabel=\theequation
              \global\@eqnswtrue
              \global\@eqcnt\z@
              \tabskip\@centering
              \let\\=\@eqncr
 \halign to \displaywidth\bgroup
    \eqnumphantom\@eqnsel\hskip\@centering
    $\displaystyle \tabskip\z@ {##}$%
    \global\@eqcnt\@ne \hskip 2\arraycolsep
         $\displaystyle\arraymode{##}$\hfil
    \global\@eqcnt\tw@ \hskip 2\arraycolsep
         $\displaystyle\tabskip\z@{##}$\hfil
         \tabskip\@centering
    &{##}\tabskip\z@\cr}
\begingroup\ifx\undefined\newsymbol \else\def\input#1 {\endgroup}\fi

\newcounter{app}

\def\app{\setcounter{equation}{0}
\def\theequation{A\Roman{app}.\arabic{equation}}\par
   \addvspace{4ex}
   \@afterindentfalse
  \secdef\@app\@dapp}
\newcommand\@app{\@startsection {app}{1}{0ex}%
                                   {-3.5ex \@plus -1ex \@minus -.2ex}%
                                   {2.3ex \@plus.2ex}%
                                   {\normalfont\Large\bf}}

\def\@dapp#1{%
{\parindent \z@ \raggedright  \bf #1}\par\nobreak}
\def\l@app#1#2{\ifnum \c@tocdepth >\z@
    \addpenalty\@secpenalty
    \addvspace{1.0em \@plus\p@}%
    \setlength\@tempdima{8.5em}%
    \begingroup
      \parindent \z@ \rightskip \@pnumwidth
      \parfillskip -\@pnumwidth
      \leavevmode \bfseries
      \advance\leftskip\@tempdima
      \hskip -\leftskip
      #1\nobreak\hfil \nobreak\hb@xt@\@pnumwidth{\hss #2}\par
    \endgroup\fi}
\newcounter{sapp}[app]

\def\sapp{\def\theequation{A\arabic{app}.\arabic{equation}}\par
   \@afterindentfalse
  \secdef\@sapp\@dsapp}
\newcommand\@sapp{\@startsection{sapp}{2}{\z@}%
                                     {-3.25ex\@plus -1ex \@minus -.2ex}%
                                     {1.5ex \@plus .2ex}%
                                     {\normalfont\large\bfseries}}

\def\@dsapp#1{%
{\parindent \z@ \raggedright  \bf #1}\par\nobreak}
\newcommand{\l@sapp}{\@dottedtocline{2}{1.5em}{3em}}
\def\draft{\oddsidemargin -.5truein
        \def\@oddfoot{\sl preliminary draft \hfil
        \rm\thepage\hfil\sl\today\quad\militarytime}
        \let\@evenfoot\@oddfoot \overfullrule 3pt
        \let\label=\draftlabel
        \let\marginnote=\draftmarginnote
   \def\@eqnnum{(\theequation)\rlap{\kern\marginparsep\tt\@eqnlabel}%
\global\let\@eqnlabel\@vacuum}  }

\def\be{\begin{eqnarray}}
\def\ee{\end{eqnarray}}

\def\p{\partial}

\def\beq{\begin{equation}}
\def\eeq{\end{equation}}
\def\ba{\beq\new\begin{array}{c}}
\def\ea{\end{array}\eeq}
\def\be{\ba}
\def\ee{\ea}

\def\Tr{{\rm Tr}\,}
\def\dim{{\rm dim}\,}

\newfont{\Bbbb}{msbm7 scaled 1\@ptsize00}

\newcommand{\z}{\raise-1pt\hbox{$\mbox{\Bbbb Z}$}}
\def\normordboson{ {\scriptstyle {{*}\atop{*}}} }

\newfont{\alef}{msbm10 at 11pt}
\newfont {\goth}{eufm10 at 11pt}
\def\mathbb#1{\hbox{{\alef #1}}}

\let\@@savethanks\thanks
\def\thanks#1{\gdef\thefootnote{\alph{footnote}}\@@savethanks{#1}}

\makeatletter
\g@addto@macro \normalsize {%
 \setlength\abovedisplayskip{14pt plus 3pt minus 3pt}%
 \setlength\belowdisplayskip{14pt plus 3pt minus 3pt}%
 \setlength\abovedisplayshortskip{11pt plus 3pt minus 3pt}%
 \setlength\belowdisplayshortskip{11pt plus 3pt minus 3pt}%
}
\makeatother

\bigskip
\bigskip

\title{
\bigskip
{\bf
Open intersection numbers and free fields} \vspace{.5cm}}
\author{{\bf Alexander Alexandrov}\thanks{E-mail:  {\tt alexandrovsash at gmail.com}}
\date{ } \\
{\small {\it 
CRM,
Universit\'e de Montr\'eal, Montr\'eal,  Canada  \&}}\\
{\small {\it 
Department of Mathematics and Statistics, Concordia University, Montreal,   Canada  \&}}\\
{\small {\it 
ITEP, Moscow, Russia}}\\
}

\begin{document}

\setcounter{footnote}{0}

\setcounter{tocdepth}{3}

\maketitle

\vspace{-8.0cm}

\begin{center}
\hfill ITEP/TH-12/16
\end{center}

\vspace{6.5cm}
\begin{abstract} 
A complete set of the Virasoro and W-constraints for the Kontsevich-Penner model, which conjecturally describes intersections on moduli spaces of open curves, was derived in our previous work. Here we show that these constraints can be described in terms of free bosonic fields with twisted boundary conditions, which gives a modification of the well-known construction of the $W^{(3)}$ algebra in conformal field theory. This description is natural from the point of view of the spectral curve description, and should serve as a new important ingredient of the topological recursion/Givental decomposition. 
\end{abstract}
\bigskip


\newpage 


\def\thefootnote{\arabic{footnote}}

\section*{Introduction}
\addcontentsline{toc}{section}{Introduction}
\setcounter{equation}{0}

Theory of open intersection numbers, which development was recently initiated in the breakthrough work of  R.~Pandharipande, J.~Solomon and R.~Tessler \cite{PST}, gives us a long-awaited open version of the Kontsevich--Witten theory \cite{Konts,Witten}. While we still don't have a comprehensive description of open intersection theory, it is clear that it fits well into the modern enumerative geometry and mathematical physics setup. Namely, the generating function of open intersection numbers can be naturally described in terms of matrix models, integrable systems, Virasoro and W-constraints, classical and quantum spectral curves, and cut-and-join operators. 

In our previous paper \cite{Aopen} we claimed that open intersection numbers can be described by the so-called Kontsevich--Penner matrix model
\be\label{inttau}
\tau_N=\det(\Lambda)^N {\mathcal C}^{-1} \int \left[d \Phi\right]\exp\left(-{\Tr\left(\frac{\Phi^3}{3!}-\frac{\Lambda^2 \Phi}{2}+N\log \Phi\right)}\right),
\ee
which belongs to the family of the generalized Kontsevich models and is a simple deformation of Kontsevich's cubic integral. The matrix of integration $\Phi$ is an $M\times M$ Hermitian matrix, it is assumed that its size $M$ tends to infinity, and the parameter of deformation $N$ is completely independent of $M$. By construction, (\ref{inttau}) is a tau-function of the modified Kadomtsev--Petviashvili (MKP) hierarchy. Using a modification of the Kac--Schwarz description for this tau-function, we derived \cite{Aopen2} a complete family of linear constraints
\be\label{intcon}
\widehat{\mathsf{L}}_{k} \tau_N =0, \,\,\,\,\,\,k\geq-1,\\
\widehat{\mathsf{M}}_{k} \tau_N =0, \,\,\,\,\,\, k\geq-2.
\ee
These constraints have a unique solution, which can be described in terms of the cut-and-join operators \cite{Aopen2}. 

However, the constraints (\ref{intcon}) are not quite satisfactory. Namely, they were constructed as a modification of the operators from $W_{1+\infty}$ algebra of symmetries of the KP hierarchy, and it is not clear why this particular modification appears. Moreover, it was not known how to represent these constraints in terms of free fields on the spectral curve. Such a representation is known for most of the Virasoro and W-constraints appearing in enumerative geometry and matrix models, and is crucially important for understanding the relations between different models. As it was shown in \cite{IMMM,IMMM1,IMMM2} this type of the representation allows us to construct the Chekhov--Eynard--Orantin (CEO) topological recursion/Givental decomposition \cite{Chekhov, Kostov, Eyn1, Eyn2, Eyn3, Giv1, Giv2}. We claim that the Kontsevich--Penner model is a fundamental ingredient of the more general version of the CEO topological recursion/Givental decomposition (to be constructed). In particular, it shall be suitable for the description of the open Gromov--Witten invariants and open topological string theory models (and, probably, for other cases, for which the quantum spectral curve degenerates into a reducible one in the classical limit). That's why it is important to have a convenient description of the linear constraints for this model.

In the current work we fill this gap and describe the Virasoro and W-constraints for open intersection numbers in terms of free bosonic fields. Obtained construction is a minor modification of the description of Zamolodchikov's \cite{Z} $W^{(3)}$ algebra given in \cite{FZ,FL}. Namely, this construction allows us to represent all generators of the $W^{(3)}$ algebra in terms of two free bosonic fields. However, unlike in the version considered in  \cite{FZ,FL}, we need to twist one of these two fields, so that it satisfies odd boundary conditions. 

The present paper is organized as follows. In Section \ref{S1} we give a brief reminder of the closed intersection theory, described by the Kontsevich--Witten tau-function. In particular, we describe the Virasoro constraints in terms of the twisted bosonic current. In Section \ref{S2} we describe some known and conjectural properties of the generating function of open intersection numbers. Section \ref{S3} is devoted to the $W_{1+\infty}$ algebra of symmetries of the KP and MKP hierarchies. In Section \ref{S4} we construct a description of the Virasoro and W-constraints for the generating function of open intersection numbers in terms of two bosonic currents. In Section \ref{S5} we show that this description is directly related to the description of the $W^{(3)}$ algebra, well known in CFT. Section \ref{S6} is devoted to the construction of the cut-and-join operators for open intersection numbers in terms of free fields.

\section{Closed intersection numbers and Kontsevich--Witten tau-function}\label{S1}

The closed intersection theory is governed by the Kontsevich--Witten tau-function. 
Let $\overline{\cal M}_{h,l}$ be the Deligne--Mumford compactification of the moduli space of genus $h$ complex curves $X$ with $l$ marked points $x_1, \dots , x_l$. 
We associate with a marked point a line bundle ${\cal L}_i$ whose fiber at a moduli point $(X;x_1,\dots,x_l)$ is the cotangent line to $X$ at $x_i$. 
 Intersection numbers of the first Chern classes of these holomorphic line bundles 
 \be\label{intclosed}
\int_{\overline{\cal M}_{h,l}}\psi_1^{\alpha_1}\psi_2^{\alpha_2}\dots\psi_l^{\alpha_l}=\langle\tau_{\alpha_1}\tau_{\alpha_2}\dots\tau_{\alpha_l}\rangle_h
\ee
are rational numbers. They are not equal to zero only if the stability condition
\be
2h-2+l>0
\ee
holds and the dimension of the moduli space
\be\label{KWdim}
\dim_{\mathbb C} {\mathcal M}_{h,l}= 3h-3+l
\ee
coincides with the degree of the form, 
\be
\dim_{\mathbb C} {\mathcal M}_{h,l}=\sum_{i-1}^l {\alpha_i}.
\label{dimcon}
\ee
Their generating function
\be\label{Ksum}
\mathcal{F}\left({\bf T},\hbar\right)=\sum_{h=0}^\infty \hbar^{2h-2}\left\langle\exp\left(\hbar\sum_{m=0}^\infty T_{m}\tau_m\right)\right\rangle_h
\ee
is given by the Kontsevich--Witten tau-function 
\be\label{KWtau}
\tau_{KW}({\bf T},\hbar)=e^{\mathcal{F}\left({\bf T},\hbar \right)}.
\ee
Let us stress that the parameter $\hbar$ is introduced in such a way that it describes not the genus expansion but the topological expansion
\be
\mathcal{F}_{KW}({\bf T},\hbar)={\sum_{\chi<0}  \hbar^{-\chi} {\mathcal F}^{(\chi)}_{KW}({\bf T})},
\ee
where
\be
\chi=2-2\# \text{handles}-\#\text{points}.
\ee
From the comparison of (\ref{KWdim}) and (\ref{dimcon}) it follows that $\hbar$ can be removed by the rescaling of $T_k$'s
\be
\tau_{KW}({\bf T},\hbar)=\tau_{KW}({\bf T},1)\Big|_{T_k\mapsto \hbar^{\frac{2k+1}{3}} T_k}.
\ee

The Kontsevich--Witten tau-function can be represented by the Kontsevich matrix integral over $M\times M$ Hermitian matrices
\be
\tau_{KW}({\bf T},\hbar)=\frac{\displaystyle{\int\left[d \Phi\right]\,\exp\left(-\frac{1}{\hbar}{\Tr\left(\frac{\Phi^3}{3!}+\frac{\Lambda \Phi^2}{2}\right)}\right)}}{\displaystyle{\int\left[d \Phi\right]\exp\left(-\frac{1}{\hbar}{\Tr\frac{\Lambda \Phi^2}{2}}\right)}}\\
= {\mathcal C}^{-1} \int \left[d \Phi\right]\exp\left(-\frac{1}{\hbar}{\Tr\left(\frac{\Phi^3}{3!}-\frac{\Lambda^2 \Phi}{2}\right)}\right),
\label{matintKon}
\ee
where
\be
{\mathcal C}=\frac{\hbar^{M^2/2}e^{\frac{1}{\hbar}\Tr \frac{\Lambda^3}{3}}}{\det\left(\Lambda^{\frac{1}{2}}\otimes 1+1\otimes \Lambda^{\frac{1}{2}}\right)}.
\ee
Asymptotic expansion of this matrix integral for the positive definite external matrix $\Lambda$ with large eigenvalues is a formal series in the variables 
 \be
T_k=(2k-1)!!\,\Tr \Lambda^{-2k-1}.
\ee
These variables can be considered as independent as the size of the matrices $M$ tends to infinity. 
 

In the rescaled time variables
\be
 t_{2k+1}=\frac{T_k}{(2k+1)!!}
\ee
the generating function (\ref{KWtau}) is a tau-function of the KdV integrable hierarchy. This tau-function satisfies the Virasoro constraints, which can be described in terms of free bosonic field with twisted boundary conditions \cite{Fuk1,DVV,Witten2}. Namely, let us consider a bosonic current 
\be
\widehat{\mathcal J}_o(x)=\frac{1}{\sqrt{2}}\sum_{k=0}^\infty\left((2k+1)\tilde{t}_{2k+1}x^{k-\frac{1}{2}}+\frac{1}{x^{k+\frac{3}{2}}}\frac{\p}{\p t_{2k+1}}\right),
\ee
where the time variables are subject to the dilaton shift
\be
\tilde{t}_{k}=t_k-\frac{\delta_{k,3}}{3\hbar}.
\ee
This current is odd when we change sheets $y\mapsto -y$ on the curve $y^2=x$. Using this current we can construct an operator
\be\label{kwviras}
\widehat{\mathcal L}(x)= \frac{1}{2}\normordboson \widehat{\mathcal J}_o^2(x)\normordboson+\frac{1}{16x^2},
\ee
where we use usual bosonic normal order which puts all $\frac{\p}{\p t_k}$ to the right of all $t_k$, and the term $\frac{1}{16x^2}$ appears because of the regularisation of the infinite sum. Then, the components of this operator
\be
\widehat{\mathcal L}(x)=\sum_{k=-\infty}^\infty \frac{\widehat{\mathcal L}_k}{x^{k+2}}
\ee
are
\be
\widehat{\mathcal L}_m=\frac{1}{4} \sum_{a+b=-1-m}(2a+1)(2 b+1) \tilde{t}_{2a+1} \tilde{t}_{2b+1}+ \frac{1}{2}\sum_{k=0}^\infty (2k+1) \tilde{t}_{2k+1} \frac{\p}{\p t_{2k+2m+1}}\\
+\frac{1}{4} \sum_{a+b=m-1} \frac{\p^2}{\p t_{2a+1} \p t_{2b+1}}+\frac{1}{16}\delta_{m,0}.
\ee
Here it is assumed that $t_k=\frac{\p}{\p t_k}=0$ for $k<0$.
These operators satisfy the Virasoro commutation relations
\be
\left[\widehat{\mathcal L}_k,\widehat{\mathcal L}_m\right]=(k-m)\widehat{\mathcal L}_{k+m}+\frac{1}{12}k(k^2-1)\delta_{k,-m}
\ee
with central charge $c=1$. A subalgebra of this Virasoro algebra annihilates the tau-function
\be\label{VirKW}
\widehat{\mathcal L}_k\, \tau_{KW}=0,\,\,\,\,\,\,\,\,\,\,\,\,\,\,\,\,\, k\geq-1.
\ee
One of the main goals of this paper is to construct a similar representation for the constraints of the generating function of open intersection numbers.

\section{Open intersection numbers and Kontsevich--Penner tau-function}\label{S2}

Of course, it is more than natural to generalize the intersection theory sketched above and to include Riemann surfaces with boundaries. However, the moduli spaces of the Riemann surfaces with boundaries are much more complicated. In particular, they are real orbifolds and so, naively they may be not-orientable. Moreover, to describe the integrals one should impose some nontrivial boundary conditions. 

That's why the moduli spaces for the open Riemann surfaces (Riemann surfaces with boundaries) were described only recently for the disc case in \cite{PST} and for the higher genera case in \cite{Tessler15}. We refer the reader to these papers for more details about this non-trivial geometric construction. However, if one includes all descendants on the boundary \cite{Buryak2}, it appears that the properties of the conjectural generating function of open intersection numbers (that is, matrix model description, integrable properties and Virasoro/W-constraints) generalize the properties of the Kontsevich--Witten tau-function in a very natural way  \cite{Aopen,Aopen2}. Here we only remind the reader some basic properties of the generating function of open intersection numbers.

\begin{figure}\label{Fig1}
\begin{center}
\begin{tikzpicture}[scale=0.9]
\filldraw[fill=lime!50,draw=white] (4,0) arc (0:360:4 and 2);
\filldraw[fill=white,draw=white] (-1.3,0) arc (-180:-360:1.3 and 0.6) arc (-30:-150:1.5 and 0.8) ;
\draw[brown!60!black] (4,0)  arc (0:360:4 and 2);
\draw  (-1.3,0) arc (-180:-360:1.3 and 0.6);
\draw  (-1.3,0) arc (-150:-180:1.5 and 0.8);
\draw  (-1.3,0) arc (-150:0:1.5 and 0.8);
\filldraw [black] (-1.9,1) circle (3pt);
\filldraw [black] (2.5,0) circle (3pt);
\begin{scope}[shift={(6,0)}]
\filldraw[fill=lime!50,draw=green!40!black] (0,0) arc (-90:90:0.5 and 1)
.. controls (3,1.2) .. (6,2)  arc (90:270:0.5 and 1)
.. controls (4.6,-1) .. (4,-3) arc (0:180:1 and 0.5)
.. controls (1.4,-1) .. (0,0);
\filldraw[fill=lime!20,draw=blue, thick] (0,0)  arc (-90:270:0.5 and 1);
\filldraw[fill=lime!20,draw=blue, thick] (6,2)  arc (-270:90:0.5 and 1);
\filldraw[fill=lime!20,draw=blue, thick] (4,-3) arc (0:360:1 and 0.5);
\filldraw[fill=white,draw=white] (1.9,0) arc (-180:-360:1.3 and 0.6) arc (-30:-150:1.5 and 0.8) ;
\draw  (1.9,0) arc (-180:-360:1.3 and 0.6);
\draw  (1.9,0) arc (-150:-180:1.5 and 0.8);
\draw  (1.9,0) arc (-150:0:1.5 and 0.8);
\filldraw [black] (2.5,-2) circle (3pt);
\filldraw [blue] (0.26,0.2) circle (3pt);
\filldraw [blue] (2.5,-3.41) circle (3pt);
\end{scope}
\end{tikzpicture}
\end{center}
\caption{Curves with $h=1,b=0,k=0, l=2$ and $h=1,b=3,k=2,l=1$.}
\end{figure}
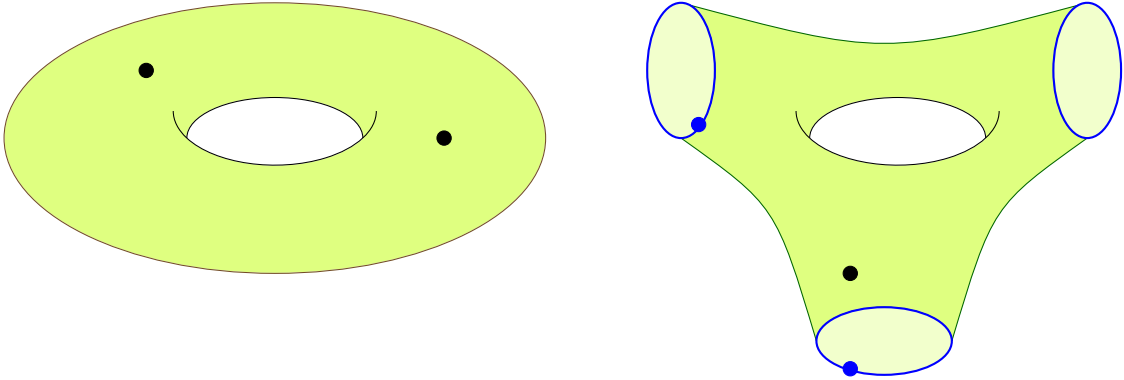

Let us consider a Riemann surface with $b$ boundaries, $h$ handles, $k$ marked points on the boundary and $l$ interior marked points. Figure 1 shows two examples: one without boundary ($b=0$) and another with three boundaries ($b=3$). In general, the corresponding moduli space has real dimension
\be\label{dimopen}
\dim_{\mathbb R} {\mathcal M}_{h,b,k,l}=6h-6+3b+k+2l,
\ee
with the stability condition
\be
4h-4+2b+k+2l>0.
\ee
In what follows we consider only the case when this dimension is even. Let us stress that we do not exclude the case with $b=0$ (without boundaries), for which the moduli spaces coincide with the closed ones considered in the previous section:
\be
{\mathcal M}_{h,0,0,l}={\mathcal M}_{h,l}.
\ee
Thus, we consider closed moduli spaces as a particular case of open ones. 

We want to find all open analogs of the intersection numbers (\ref{intclosed}), that is
\be\label{gencor}
``\int_{\overline{\mathcal M}_{h,b,k,l}}\psi_1^{\alpha_1}\dots\psi_l^{\alpha_l}{\phi}_{l+1}^{\beta_1}\dots{\phi}_{l+k}^{\beta_k}"=\left<\tau_{\alpha_1}\dots\tau_{\alpha_l}\sigma_{\beta_1}\dots\sigma_{\beta_k}\right>_{h,b},
\ee
where $\psi_j$ are the first Chern classes of the bundles ${\cal L}_i$ corresponding to the interior marked points and ${\phi}_{j}$ are their analogs for the boundary marked points. This intersection number is not equal to zero only if the dimension condition is satisfied:
\be\label{dimopen2}
\dim_{\mathbb R} {\mathcal M}_{h,b,k,l}=2\sum_{i=1}^l \alpha_i +2\sum_{i=1}^k\beta_i. 
\ee
One can define the generating function of all these intersection numbers
\be\label{Fopen}
\mathcal{F}_Q({\bf T};{\bf S},\hbar)=\sum_{h=0}^{\infty}\sum_{b=0}^{\infty} \hbar^{2h-2+b} Q^{b}
\left<\exp\left(\hbar \sum_{k \geq 0} (T_{k} \tau_k+  S_{k}\sigma_k) \right)\right>_{h,b}
\ee
and
\be\label{Qtau}
\tau_{Q}({\bf T};{\bf S}, \hbar)=e^{\mathcal{F}_Q({\bf T};{\bf S},\hbar)}.
\ee
$\hbar$-expansion gives a topological expansion 
\be
\mathcal{F}_Q({\bf T};{\bf S},\hbar)={\sum_{\chi<0}  \hbar^{-\chi} {\mathcal F}^{(\chi)}_Q({\bf T};{\bf S})},
\ee
where
\be
\chi=2-2\# \text{handles}-\# \text{boundaries}-\#\text{points}.
\ee
From the dimension constraints (\ref{dimopen}) and (\ref{dimopen2}) it follows that the power of $\hbar $ in the sum (\ref{Fopen}) in front of the correlation function (\ref{gencor}) is
\be\label{dimope}
2h-2+b+k+l=\frac{1}{3}\left(\sum_{i=1}^l (2\alpha_i+1)+\sum_{i=1}^k(2\beta_i+2) \right),
\ee
so that
\be
\tau_{Q}({\bf T};{\bf S}, \hbar)=\tau_{Q}({\bf T};{\bf S}, 1)\Big|_{T_k\mapsto \hbar^{\frac{2k+1}{3}} T_k, \,S_k\mapsto \hbar^{\frac{2k+2}{3}} S_k}.
\ee
In what follows we omit the $\hbar$ dependence and restore it only in Section \ref{S6}, when we consider a topological recursion.

From the definition (\ref{Fopen}) it follows that for $Q=0$ only the components without boundaries contribute to the sum, so that the generating function does not depend on $S_k$'s and coincides with the Kontsevich--Witten tau-function
\be
\tau_{0}({\bf T};{\bf S})=\tau_{KW}({\bf T}).
\ee

We use quotation marks in (\ref{gencor}) because a complete definition of the classes ${\phi}_j$ is still not available and, moreover, even for the available cases to specify the boundary conditions one should introduce more involved ingredients, in particular the canonical multisections. However, in \cite{PST} all the intersection numbers of the form
\be
``\int_{\overline{\mathcal M}_{0,1,k,l}}\psi_1^{\alpha_1}\dots\psi_l^{\alpha_l}{\phi}_{l+1}^{0}\dots{\phi}_{l+k}^{0}"
\ee
were computed.  The authors also gave a conjectural description of the generating function 
\be
\tau_{1}({\bf T};{\bf S_0}). 
\ee
This function corresponds to the specification of (\ref{Qtau}) with $Q=1$ (the case, when the components of the moduli spaces with different number of boundaries contribute with the same weight) and ${\bf S_0}=\{S_0,0,0,\dots\}$ (without descendants on the boundary).

Namely, the authors of \cite{PST} conjectured that this generating function satisfies some analog of the KdV equations (called there open KdV equations) and some Virasoro constraints (which, in particular, are not quadratic in derivatives and contain all higher derivatives with respect to $S_0$). These conjectures were proved there for $h=0, b=1$.

In the works of Buryak \cite{Buryak, Buryak2} it was shown that the open KdV equations and the Virasoro constraints of \cite{PST} are consistent and equivalent. Moreover, a very explicit and simple conjectural formula for the generating function
\be\label{Burgf}
\tau_{1}({\bf T};{\bf S})
\ee
was suggested. It was shown that this generating function satisfies the Virasoro constraints (which include derivatives of degree not higher when two) and can be considered as a wave function of the KdV hierarchy.

On the basis of the works \cite{PST, Buryak, Buryak2} the author suggested in \cite{Aopen,Aopen2} a matrix model description of the generating function of open intersection numbers. Namely, it was conjectured in \cite{Aopen} and later proved in \cite{Aopen2} that the so-called Kontsevich--Penner matrix integral
\be
\tau_N=\frac{\displaystyle{\int\left[d \Phi\right]\,\det\left(1+\frac{\Phi}{\Lambda}\right)^{-N}\exp\left(-{\Tr\left(\frac{\Phi^3}{3!}+\frac{\Lambda \Phi^2}{2}\right)}\right)}}{\displaystyle{\int\left[d \Phi\right]\exp\left(-{\Tr\frac{\Lambda \Phi^2}{2}}\right)}}\\
=\det(\Lambda)^N {\mathcal C}^{-1} \int \left[d \Phi\right]\exp\left(-{\Tr\left(\frac{\Phi^3}{3!}-\frac{\Lambda^2 \Phi}{2}+N\log \Phi\right)}\right),
\label{matint}
\ee
which obviously  coincides with the Kontsevich matrix integral (\ref{matintKon}) for $N=0$, for $N=1$ gives the expansion of (\ref{Burgf}). A parameter $N$ is assumed to be independent of the size of the matrices $M$ and
\be
T_k=(2k-1)!!\,\Tr \Lambda^{-2k-1},\\
S_k=2^{k} k! \, \Tr \Lambda^{-2k-2}.
\ee
Matrix model (\ref{matint}) belongs to the family of the so called generalised Kontsevich models and was considered earliar in \cite{BH}. From this matrix integral expression it immediately follows that (\ref{Burgf}) is a tau-function of the KP integrable hierarchy in the KP times $t_k$,
\be
t_{2k+1}=\frac{T_k}{(2k+1)!!},\\
t_{2k+2}=\frac{S_k}{2^{k+1} (k+1)!} \, .
\ee

The Virasoro constraints satisfied by (\ref{matint}) for $N=1$ (actually, a one-parametric family of the Virasoro constraints which includes, in particular, constraints from  \cite{Buryak}) were also constructed in \cite{Aopen}. The main difference with the Virasoro constraints of \cite{PST, Buryak, Buryak2} is that a representative of this family of the Virasoro constraints belong to the $W_{1+\infty}$ symmetry algebra of the KP integrable hierarchy and are naturally related to the stabilizer of the corresponding point of the Sato Grassmannian. 

While the identification between the generating function (\ref{Qtau}) and the tau-function (\ref{matint}) for $N=Q=0$ has been known for more whan 25 years, and for $N=Q=1$ was established in \cite{Aopen,Aopen2}, only  Safnuk in \cite{Safnuk} conjectured that the Kontsevich--Penner matrix model actually describes a refined version of intersection theory, and the parameter $N$ in (\ref{matint}) should be identified with $Q$ in (\ref{Qtau}). It can be shown that this conjecture is true \cite{ABT}. That's why it is important to consider the model (\ref{matint}) for arbitrary values of $N$ and, in particular, to understand the structure of the equations satisfied by this model.

Generalized Kontsevich models are the tau-functions of the KP/Toda type hierarchies\cite{Toda}. In particular, matrix model (\ref{matint}) defines a tau-function of the MKP hierarchy dependent on the parameter $N$ so that (\ref{Burgf}) and (\ref{KWtau}) are related to each other by the MKP Hirota bilinear identity. So far it is not clear if the open KdV hierarchy of \cite{PST} is of any fundamental importance. In particular, we don't know if there are any other interesting solutions (for example, solitons) or if this hierarchy can be represented in the bilinear Hirota form. That's why we claim that namely MKP is an integrable structure which governs open intersection numbers, and open KdV is only a consequence of MKP and other relations valid for $N=1$ but not for other values of $N$. Full power of MKP is seen only if we consider the dependence (\ref{matint}) of all time variables $t_k$ and $N$.  

Virasoro and W-constraints for an arbitrary $N$ were constructed in \cite{Aopen2} with the help of the modification of the matrix models and integrable hierarchies methods. For $N=0$ and $N=1$ the constraints belong to the $W_{1+\infty}$ algebra of symmetries of the KP integrable hierarchy, for other values of $N$ this is not true anymore. In spite of the existence of explicit formulas the structure of the constraints was not clear so far.

Below we explain the structure of the Virasoro and W-constraints satisfied by (\ref{matint}). Namely, we show that these constraints can be naturally described in terms of a subalgebra of the $W^{(3)}$-algebra given by a simple modification of the construction well known in CFT. 

\section{$W_{1+\infty}$ algebra and (M)KP}\label{S3}

From the free fermion description of the KP hierarchy it immediately follows that the operators\footnote{For more details see, i.e., \cite{Sato,Segal,Fukuma, AHeis} and references therein.}
\be
\widehat{W}^{(m+1)}(z)=\normordboson 
\left(\widehat{J}(z)+\p_z \right)^m \widehat{J}(z)\normordboson
\ee
correspond to the bilinear combinations of fermions and span the  $W_{1+\infty}$ algebra of symmetries of the KP hierarchy.
Here $\widehat{J}(z)$ is the bosonic current
\be\label{boscurKP}
\widehat{J}(z)= \sum_{m \in \z} \frac{\widehat{J}_m}{z^{m+1}},
\ee
where
\be\label{KPJ}
\widehat{J}_k =
\begin{cases}
\displaystyle{\frac{\p}{\p t_k} \,\,\,\,\,\,\,\,\,\,\,\, \mathrm{for} \quad k>0},\\[10pt]
\displaystyle{0}\,\,\,\,\,\,\,\,\,\,\,\,\,\,\,\,\,\,\, \mathrm{for} \quad k=0,\\[10pt]
\displaystyle{-kt_{-k} \,\,\,\,\,\mathrm{for} \quad k<0.}
\end{cases}
\ee
The normal ordering for bosonic operators $\normordboson\dots\normordboson$ puts all operators $\widehat{J}_k$ with positive $k$ to the right of all $\widehat{J}_k$ with negative $k$. In what follows we need only operators with $m=0,1,2$.

The Virasoro subalgebra of $W_{1+\infty}$ is generated by the operators, bilinear in $\widehat{J}_k$,
\be
\frac{1}{2}\normordboson \widehat{J}(z)^2\normordboson = \sum_{m\in \z} \frac{\widehat{L}_m}{z^{m+2}},
\ee
namely it is spanned by the operators
\begin{equation}
\label{virfull}
\widehat{L}_m=\frac{1}{2} \sum_{a+b=-m}a b t_a t_b+ \sum_{k=1}^\infty k t_k \frac{\p}{\p t_{k+m}}+\frac{1}{2} \sum_{a+b=m} \frac{\p^2}{\p t_a \p t_b}.
\end{equation}
The operators from $W_{1+\infty}$, cubic in $\widehat{J}_k$,
\begin{multline}
\widehat{M}_k=\frac{1}{3} \sum_{a+b+c=k} \normordboson \widehat{J}_a \widehat{J}_b \widehat{J}_c \normordboson=
\frac{1}{3}\sum_{a+b+c=-k}a\, b\, c\, t_a\, t_b\, t_c+\sum_{c-a-b=k}a\, b\, t_a\, t_b\, \frac{\p}{\p t_{c}}\\
+\sum_{b+c-a=k}a\, t_{a}\frac{\p^2}{\p t_b\p t_c}+\frac{1}{3}\sum_{a+b+c=k}\frac{\p^3}{\p t_a \p t_b \p t_c},
\end{multline}
are generated by
\be
\frac{1}{3}\normordboson \widehat{J}(z)^3\normordboson=\sum_{m\in \z}\frac{\widehat{M}_m}{z^{m+3}}.
\ee
The operators $\widehat{J}_k$, $\widehat{L}_k$, and $\widehat{M}_k$ satisfy the following commutation relations
\be\label{Wcomr}
\left[\widehat{J}_k,\widehat{J}_m\right]=k\, \delta_{k,-m},\\
\left[\widehat{J}_k,\widehat{L}_m\right]=k\widehat{J}_{k+m},\\
\left[\widehat{L}_k,\widehat{L}_m\right]=(k-m)\widehat{L}_{k+m}+\frac{1}{12}k(k^2-1)\delta_{k,-m},\\
\left[\widehat{L}_k,\widehat{M}_m\right]=(2k-m)\widehat{M}_{k+m}+\frac{1}{6}k(k^2-1)\widehat{J}_{k+m},\\
\left[\widehat{J}_k,\widehat{M}_m\right]=2k\,\widehat{L}_{k+m}.\\
\ee
A commutator of $\widehat{M}_k$'s contains the terms of fourth power of the current components $\widehat{J}_m$, so it can not be represented as a linear combination of $\widehat{J}_k$, $\widehat{L}_k$, and $\widehat{M}_k$.

Constraints (\ref{VirKW}) for the Kontsevich--Witten tau-function can be represented in terms of the $W_{1+\infty}$ algebra of symmetries of the KP hierarchy. Namely, the reduction to the KdV hierarchy is described by
\be
\widehat{J}_{2k}\, \tau_{KW}=0, \,\,\,\,\,\,\, \quad k\geq 1,
\ee
and the Virasoro constraints are given by
\be
\left(\frac{1}{2}\widehat{L}_{2k}-\frac{1}{2}\widehat{J}_{2k+3}+\frac{1}{16}\delta_{k,0}\right)\tau_{KW}=0,~~~~~k\geq -1.
\ee

To describe the symmetries of the MKP hierarchy one have to introduce an additional variable $N$ and
consider a generalization of the above description. 
Namely, instead of (\ref{KPJ}) we have 
\be
\widehat{J}^{MKP}_k =
\begin{cases}
\displaystyle{\frac{\p}{\p t_k} \,\,\,\,\,\,\,\,\,\,\,\, \mathrm{for} \quad k>0},\\[10pt]
\displaystyle{N}\,\,\,\,\,\,\,\,\,\,\,\,\,\,\,\,\,\,\, \mathrm{for} \quad k=0,\\[10pt]
\displaystyle{-kt_{-k} \,\,\,\,\,\mathrm{for} \quad k<0.}
\end{cases}
\ee
Then the symmetry operators of the MKP hiearchy, given by the coefficients of 
\be\label{JMKP}
\widehat{J}^{MKP}(z)= \sum_{m \in \z} \frac{\widehat{J}^{MKP}_m}{z^{m+1}},
\ee 
$\frac{1}{2}\normordboson \widehat{J}^{MKP}(z)^2\normordboson$ and $\frac{1}{3}\normordboson \widehat{J}^{MKP}(z)^3\normordboson$ are
 \be\label{MKPop}
\widehat J^{MKP}_k=\widehat J_k+\delta_{k,0} N,\\
\widehat L^{MKP}_k=\widehat L_k+N\widehat J_k+\frac{N^2}{2}\delta_{k,0},\\
\widehat M^{MKP}_k= \widehat M_k +2N \widehat L_k +N^2 \widehat J_k +\frac{N^3}{3}\delta_{k,0}.
\ee
They satisfy the same commutation relations (\ref{Wcomr}). Symmetry operators for the KP hierarchy can be expressed as a linear combination of the operators for the MKP hierarchy:
\be
\widehat J_k=\widehat J^{MKP}_k-\delta_{k,0} N,\\
\widehat L_k=\widehat L^{MKP}_k-N\widehat J^{MKP}_k+\frac{N^2}{2}\delta_{k,0},\\
\widehat M_k= \widehat M^{MKP}_k -2N \widehat L^{MKP}_k +N^2 \widehat J^{MKP}_k -\frac{N^3}{3}\delta_{k,0}.
\ee

\section{Constrains for open intersection numbers}\label{S4}

In \cite{Aopen2} it is proven that the Kontsevich-Penner tau-function (\ref{matint}) satisfies the linear equations
\be
\widehat{\mathsf{L}}_{k} \tau_N =0, \,\,\,\,\,\,k\geq-1,\\
\widehat{\mathsf{M}}_{k} \tau_N =0, \,\,\,\,\,\, k\geq-2.
\ee
Here
 \be\label{Virall}
\widehat{\mathsf{L}}_{k}=\widehat{L}_{2k}-\frac{\p}{\p t_{2k+3}}+3N\frac{\p}{\p t_{2k}}+\sum_{j=1}^{k-1}\frac{\p^2}{\p t_{2j}\p t_{2k-2j}}+\left(\frac{1}{8}+\frac{3N^2}{2}\right)\delta_{k,0}+2Nt_2\delta_{k,-1},
 \ee
 and
\be\label{W3all}
\widehat{\mathsf{M}}_{k}=\widehat{M}_{2k}-2\widehat{L}_{2k+3}+\widehat{J}_{2k+6}+\left(3(k+1)N^2+\frac{1}{4}\right)\widehat{J}_{2k}\\
+(k+4)N\left(\widehat{L}_{2k}-\widehat{J}_{2k+3}\right)+2\left(N^2+\frac{1}{4}\right)N\delta_{k,0}+4\,N^2t_2\delta_{k,-1}+16\,N^2t_4\delta_{k,-2}\\
+(k-2)N\sum_{j=1}^{k-1}\frac{\p^2}{\p t_{2j}\p t_{2k-2j}}-\frac{4}{3}\sum_{i+j+l=k}\frac{\p^3}{\p t_{2i}\p t_{2j}\p t_{2l}}.
\ee
Rather involved commutation relations
\be
\left[\widehat{\mathsf{L}}_{k},\widehat{\mathsf{L}}_{m}\right]=2(k-m)\widehat{\mathsf{L}}_{k+m},\\
\left[\widehat{\mathsf{L}}_{k},\widehat{\mathsf{M}}_{l}\right]=2\,(2k-l)\widehat{\mathsf{M}}_{k+l}-4\left(k(k-1)-2\delta_{k,-1}\right)\,N\,\widehat{\mathsf{L}}_{k+l}+8\sum_{j=1}^{k-1}j \frac{\p}{\p t_{2k-j}}\widehat{\mathsf{L}}_{l+j},
\ee
indicate that the basis of constraints (\ref{Virall}), (\ref{W3all}) is not the most convenient one. Below we obtain a more natural basis of constraints, with simple commutation relations and a natural free field representation.

From (\ref{Virall}) and (\ref{W3all}) it is clear that these constraints can not be represented in terms of the bosonic current (\ref{boscurKP}) since the symmetry between even and odd times is broken. Thus, let us consider even and odd times separately, and introduce two components of the bosonic current
\be\label{eocurrents}
\widehat J_o(z)=v_o(z)+\widehat{\nabla}_o(z),\\
\widehat J_e(z)=v_e(z)+\widehat{\nabla}_e(z),
\ee
where
\be
v_o(z)=\sum_{k=0}^{\infty} {(2k+1) z^{2k}\tilde{t}_{2k+1}},\\
\widehat{\nabla}_o(z)=\sum_{k=0}^\infty \frac{1}{z^{2k+2}}\frac{\p}{\p t_{2k+1}},
\ee
and
\be
v_e(z)=\sum_{k=1}^{\infty} 2k z^{2k-1} {t}_{2k},\\
\widehat{\nabla}_e(z)=\frac{N}{z}+\sum_{k=1}^\infty \frac{1}{z^{2k+1}}\frac{\p}{\p t_{2k}}.
\ee
Here the times $\tilde{t}_k$ are subject to the dilaton shift
\be\label{dilsh}
\tilde{t}_k=t_k-\frac{\delta_{k,3}}{3}.
\ee
From (\ref{JMKP}) it is easy to see that
\be
\widehat J_o(z)+\widehat J_e(z)=\widehat J^{MKP}(z)-z^2.
\ee


We introduce new operators
\be
\widehat{\mathsf{M}}_k':=\widehat{\mathsf{M}}_{k}-(k+2)N\widehat{\mathsf{L}}_{k},
\ee
such that $\widehat{\mathsf{M}}_k'$ and $\widehat{\mathsf{L}}_{k}$ constitute a basis of constraints equivalent to the basis (\ref{Virall}) and (\ref{W3all}). These new operators can be represented as a linear combination of the operators (\ref{MKPop}) and the powers of $\widehat\nabla_e(z)$:
\be\label{newop}
\widehat{\mathsf{L}}_{k}=\widehat{L}^{MKP}_{2k}-\widehat{J}^{MKP}_{2k-3} +\frac{\delta_{k,0}}{8}+\frac{1}{2\pi i}\oint \widehat\nabla_e(z)^2 z^{2k+1}dz,\\
\widehat{\mathsf{M}}_k'
=\widehat{{M}}^{MKP}_{2k}-2\widehat{{L}}^{MKP}_{2k-3}+\widehat{{J}}^{MKP}_{2k-6}+\frac{1}{4}\widehat{{J}}^{MKP}_{2k}-\frac{2}{3\pi i} \oint \widehat\nabla_e(z)^3 z^{2k+2}dz.
\ee
In this new basis the commutation relations are
\be
\left[\widehat{\mathsf{L}}_{k},\widehat{\mathsf{M}}_l'\right]=2(2k+l)\widehat{\mathsf{M}}_{k+l}'+8\sum_{j=1}^k j \widehat{J}^{MKP}_{2k-2j}\widehat{\mathsf{L}}_{l+j}.
\ee

Then the generating functions of the operators (\ref{newop}) are
\be\label{}
\sum_{k=-1}^\infty \frac{\widehat{\mathsf{L}}_{k}}{z^{2k+2}}=\frac{1}{2} z^2\left(z^{-2}\normordboson  \widehat{ J}_o(z)^2 +\widehat J_e(z)^2+2 \widehat\nabla_e(z)^2+\frac{1}{4z^2} \normordboson\right)_-,\\
\sum_{k=-2}^\infty \frac{\widehat{\mathsf{M}}_k'}{z^{2k+3}}=\frac{1}{3} z^2\left(z^{-2}\normordboson  3\widehat{ J}_o(z)^2\widehat J_e(z)+\widehat J_e(z)^3 -4 \widehat\nabla_e(z)^3+\frac{3}{4z^2}\widehat J_e(z)  \normordboson\right)_-
\ee
where we use the standard notation for the negative part of the Laurent series
\be
\left(\sum_{k=-\infty}^\infty a_k x^k\right)_-:=\sum_{k=-\infty}^{-1} a_k x^k.
\ee

However, this representation still contains not only the currents, but also the operator $\widehat\nabla_e(z)$, so it is not quite satisfactory. Let us introduce new operators $\widehat{\mathsf{M}}_k^*$ using the generating function:
\be
\sum_{k=-2}^\infty \frac{\widehat{\mathsf{M}}_k^*}{z^{2k+3}}:=\sum_{k=-2}^\infty \frac{\widehat{\mathsf{M}}_k'}{z^{2k+3}} - \frac{4}{3}z^2\left(z^{-2}v_0(z)\sum_{k=-1}^\infty \frac{\widehat{\mathsf{L}}_{k}}{z^{2k+2}}\right)_-\\
=\frac{1}{3}z^2 \left(z^{-2}\normordboson  3\widehat{ J}_o(z)^2\widehat J_e(z)+\widehat J_e(z)^3 -4 \widehat{\nabla}_e(z)^3+\frac{3}{4z^2}\widehat J_e(z)  \normordboson\right)_-\\
 - \frac{2}{3}z^{2}\left(z^{-2}\normordboson  v_e(z)\widehat{ J}_o(z)^2 +v_e(z)\widehat J_e(z)^2+2v_e(z) \nabla_e(z)^2 \normordboson+\frac{v_e(z)}{4z^2} \right)_-\\
 =z^{2}\left(z^{-2}\normordboson \left(\frac{1}{3}v_e(z)+\widehat\nabla_e(z)\right)\left(\widehat { J}_o(z)^2+\frac{1}{4z^2}\right)-\left(\frac{1}{3}v_e(z)+\widehat\nabla_e(z)\right)^3  \normordboson\right)_-
\ee
which, by construction, also annihilate the Kontsevich--Penner tau-function. It is easy to show that now the canonical commutation relations are satisfied
\be
\left[\widehat{\mathsf{L}}_{k},\widehat{\mathsf{M}}_k^*\right]=2(2k-l)\widehat{\mathsf{M}}_{k+l}^*.
\ee

To describe the operators $\widehat{\mathsf{M}}_k^*$ and $\widehat{\mathsf{L}}_{k}$ in terms of free fields let us introduce
\be
\widehat{ \mathcal J}_o(x)=\frac{1}{\sqrt{2}}\sum_{k=0}^\infty\left((2k+1)\tilde{t}_{2k+1}x^{k-\frac{1}{2}}+\frac{1}{x^{k+\frac{3}{2}}}\frac{\p}{\p t_{2k+1}}\right),\\
\widehat{ \mathcal J}_e(x)=\sum_{k=0}^\infty\left(\sqrt{\frac{2}{3}}{k}\,\tilde{t}_{2k}x^{k-1}+\sqrt{\frac{3}{2}}\frac{1}{x^{k+1}}\frac{\p}{\p t_{2k}}\right)+\sqrt{\frac{3}{2}}\frac{N}{x},
\ee
where the time variables are subject to the dilaton shift (\ref{dilsh}).
We see that the odd current $\widehat{ \mathcal J}_o(z)$ coincides with the current from the description of the Kontsevich-Witten tau-function in Section \ref{S1}.
Then, if we consider
\be\label{newnewbasis}
\widehat{\mathcal{L}}^N(x)=\sum_{k=-\infty}^\infty \frac{\widehat{\mathcal{L}}_{k}^N}{x^{k+2}}:=\frac{1}{2}\left(\normordboson  \widehat{{ \mathcal J}}_o(x)^2+\frac{1}{8x^2} +\widehat { { \mathcal J}}_e(x)^2 \normordboson\right),\\
\widehat{\mathcal{M}}^N(x)=\sum_{k=-\infty}^\infty \frac{\widehat{\mathcal{M}}_{k}^N}{x^{k+3}}:=\frac{1}{\sqrt{6}}\left(\normordboson \widehat {{ \mathcal J}}_e(x)\left(\widehat{ { \mathcal J}}_o(x)^2+\frac{1}{8x^2}\right)-\frac{1}{3}\widehat { { \mathcal J}}_e(x)^3  \normordboson\right),
\ee
the operators $\widehat{\mathcal{L}}_{k}^N$ and $\widehat{\mathcal{M}}_{k}^N$ coincide, up to normalisation, with the operators $\widehat{\mathsf{L}}_{k}$ and $\widehat{\mathsf{M}}_k^*$, and satisfy the commutation relations
\be\label{mastcr}
\left[\widehat{\mathcal L}_k^N,\widehat{\mathcal L}_m^N\right]=(k-m)\widehat{\mathcal L}_{k+m}^N+\frac{1}{6}k(k^2-1)\delta_{k,-m},\\
\left[\widehat{\mathcal L}_k^N,\widehat{\mathcal M}_m^N\right]=(2k-m)\widehat{\mathcal M}_{k+m}^N,\\
\left[\widehat{\mathcal M}_k^N,\widehat{\mathcal M}_m^N\right]=\frac{1}{2}(k-m)\widehat{\Lambda}_{k+m}^N+\frac{1}{180}k(k^2-1)(k^2-4)\delta_{k,-m}\\
+(k-m)\left(\frac{1}{15}(k+m+3)(k+m+2)-\frac{1}{6}(k+2)(m+2)\right)\widehat{\mathcal L}_{k+m}^N,
\ee
where
\be
\widehat{\Lambda}^N_m=\sum_{n\leq -2} \widehat{\mathcal L}_n^N \widehat{\mathcal L}_{m-n}^N+\sum_{n>2} \widehat{\mathcal L}_{m-n}^N \widehat{\mathcal L}_{n}^N-\frac{3}{10}(m+3)(m+2)\widehat{\mathcal L}_m^N.
\ee
These operators generate a representation of the $W^{(3)}$ algebra with central charge $c=2$. 
Then the constraints for the Kontsevich--Penner model are given by
\be\label{CONS}
\left(\widehat{\mathcal{L}}^N(x)\right)_-\tau_N=\left(\widehat{\mathcal{M}}^N(x)\right)_-\tau_N=0.
\ee

Equations (\ref{CONS}) can be considered as the Virasoro and W-master equations \cite{KaZj}. They can serve as a basis for the derivation of the CEO topological relations/Givental decomposition for the ramified generating function of open intersection numbers \cite{Afrie}.

\section{$W^{(3)}$-algebra and free bosonic fields}\label{S5}

Representation (\ref{newnewbasis}) of the $W^{(3)}$ algebra can naturally be described by the construction, well known in the conformal field theory \cite{FZ,FL} and in matrix models \cite{Fukuma,Confmm0,Confmm}. This construction is related to the Miura transformation. 

Let us describe this construction in a bit more general setup. For the case of $sl(n)$ the $W^{(n)}$ algebra can be represented in terms of the vector of $n-1$ independent bosonic currents $\vec{J}=(J_{(1)}(x), J_{(2)}(x),\dots, J_{(n-1)}(x) )$ 
\be
J_{(k)}(x)=\p_x \phi_{(k)}(x)=\sum_{m=-\infty}^\infty \frac{J^{(k)}_m}{ x^{m+1}},
\ee
where operators $J^{(k)}_m$ satisfy the commutation relations
\be
\left[J^{(k)}_m,J^{(l)}_n\right]=m\,\delta_{k,l}\,\delta_{m,-n}.
\ee

Let us consider a generating function of the operators $W^{(j)}$ (dependent of $x$)
\be
R_n(u)=-\normordboson\prod_{m=1}^n(u-\vec{h}_m\cdot\vec{J})\normordboson,
\ee
which is polynomial in the new variable $u$:
\be
R_n(u)=-u^n+\sum_{j=2}^n u^{n-j} W^{(j)}.
\ee
Here the $\vec{h}_m$'s are the weight vectors of the fundamental representation of $sl(n)$ and they satisfy
\be
\vec{h}_i\cdot\vec{h}_j=\delta_{ij}-\frac{1}{n},\,\,\,\,\,\,\,\,\,\, \sum_{j=1}^n \vec{h}_j=0.
\ee
Then the operators $W^{(j)}$ generate a representation of the $W^{(n)}$ algebra with central charge $c=n-1$.

In particular, for $n=3$ we have
\be
\vec{h}_1=\left(\frac{1}{\sqrt{2}},\frac{1}{\sqrt{6}}\right),\,\,\,\,\,\vec{h}_2=\left(-\frac{1}{\sqrt{2}},\frac{1}{\sqrt{6}}\right),\,\,\,\,\,\vec{h}_3=\left(0,-\frac{2}{\sqrt{6}}\right),
\ee
and the $W^{(3)}$ algebra with $c=2$ is generated by
\begin{equation}
\begin{gathered}
R_3(u)=-\normordboson\prod_{m=1}^3(u-\vec{h}_m \vec{J})\normordboson\\
=-u^3-u\normordboson\prod_{i<j}(\vec{h}_i \cdot\vec{J})(\vec{h}_j\cdot\vec{J})\normordboson+\normordboson\prod_{i}\vec{h}_i\cdot\vec{J}\normordboson\\=
-u^3+u\,{\mathcal{L}}(x)+{\mathcal{M}}(x).
\end{gathered}
\end{equation}
Here
\begin{equation}\label{CFTop}
\begin{gathered}
{\mathcal{L}}(x)=\sum_{k=\infty}^\infty \frac{{\mathcal{L}}_{k}}{x^{k+2}}=\frac{1}{2}\left(\normordboson  {J}_{(1)}(x)^2 + { J}_{(2)}(x)^2 \normordboson\right),\\
{\mathcal{M}}(x)=\sum_{k=-\infty}^\infty \frac{{\mathcal{M}}_{k}}{x^{k+3}}:=\frac{1}{\sqrt{6}}\left(\normordboson  { J}_{(1)}(x)^2  {J}_{(2)}(x)-\frac{1}{3} { J}_{(2)}(x)^3  \normordboson\right).
\end{gathered}
\end{equation}

We see that these expressions almost coincide with the expressions for $\widehat{\mathcal{L}}^N(x)$ and $\widehat{\mathcal{M}}^N(x)$ (\ref{newnewbasis}). Namely, to get (\ref{newnewbasis}) we have to substitute 
\be
{ J}_{(1)}(x)^2 \mapsto \widehat{ { \mathcal J}}_o(x)^2+\frac{1}{8x^2},\\ 
{J}_{(2)}(x) \mapsto \widehat{ { \mathcal J}}_e(x).
\ee
Clearly, this substitution does not affect the commutation relations. Thus, operators generated by (\ref{newnewbasis}) and (\ref{CFTop}) satisfy the same commutation relations (\ref{mastcr}). 

Therefore, we see that the constraint for the generating function of open intersection numbers can be obtained by a slight modification of the construction well known in the CFT. The modification corresponds to the twist of one (namely $\phi_{(1)}$) of the bosonic fields.\footnote{If we twist both fields with $Z_3$-twist, we will obtain the constraints for the tau-function of the Boussinesq hierarchy, corresponding to the generalized Kontsevich model with quartic potential \cite{Fukuma}.} Corresponding current $J_{(1)}$ should satisfy odd boundary conditions to describe the constraints for open intersection numbers. Unfortunately, we were unable to find this representation of the $W^{(3)}$ algebra in the literature. 

This description of the constraints can be easily generalized for the case of r-spin open intersection numbers \cite{AMKP}.

\section{Cut-and-join description and topological recursion}\label{S6}

Let us describe a topological recursion which gives a solution of the constraints (\ref{CONS}). From (\ref{dimope}) it follows that
\be
\hbar \frac{\p}{\p \hbar}\, \tau_{N}({\bf t},\hbar)=\widehat{D} \, \tau_{N}({\bf t},\hbar)
\ee
with the degree operator
\be
\widehat{D}:=\frac{1}{3}\sum_{k=1}^\infty k t_k \frac{\p}{\p t_k}.
\ee
We can rewrite this operator as
\be
\widehat{D} =\frac{1}{6\pi i}\oint \left(\sqrt{\frac{2}{3}}v_e(\sqrt{x})\widehat { { \mathcal J}}_e(x)+\sqrt{2}v_o(\sqrt{x})\widehat { { \mathcal J}}_o'(x)\right)\sqrt{x} dx,
\ee
where
\be
\widehat{ \mathcal J}_o'(x)=\frac{1}{\sqrt{2}}\sum_{k=0}^\infty\left((2k+1){t}_{2k+1}x^{k-\frac{1}{2}}+\frac{1}{x^{k+\frac{3}{2}}}\frac{\p}{\p t_{2k+1}}\right)=\widehat{ \mathcal J}_o(x)+\sqrt{\frac{x}{2}}
\ee
is the odd current without dilaton shift. Then, if we rewrite the operators (\ref{newnewbasis}) as
\be
\widehat{\mathcal{L}}^N(x)=\frac{1}{2}\left(\normordboson \left( \widehat{{ \mathcal J}}_o'(x)-\sqrt{\frac{x}{2}}\right)^2+\frac{1}{8x^2} +\widehat { { \mathcal J}}_e(x)^2 \normordboson\right),\\
\widehat{\mathcal{M}}^N(x)=\frac{1}{\sqrt{6}}\left(\normordboson \widehat {{ \mathcal J}}_e(x)\left(\left( \widehat{{ \mathcal J}}_o'(x)-\sqrt{\frac{x}{2}}\right)^2+\frac{1}{8x^2}\right)-\frac{1}{3}\widehat { { \mathcal J}}_e(x)^3  \normordboson\right),
\ee
from the constraints (\ref{CONS}) it follows that $\tau_N({\bf t},\hbar)$ satisfies the cut-and-join type equation
\be
\widehat{D} \tau_N({\bf t},\hbar)=\left(\hbar\, \widehat{\mathcal{W}}_{1}+\hbar^2\widehat{\mathcal{W}}_{2}\right) \tau_N({\bf t},\hbar),
\ee
where
\be
\widehat{\mathcal{W}}_1=\frac{1}{6\pi i} \oint\left(\normordboson v_o(\sqrt{x})\left( \widehat{{ \mathcal J}}_o'(x)^2+\frac{1}{8x^2}+\widehat { { \mathcal J}}_e(x)^2\right)+\frac{4v_e(z)}{\sqrt{3}}\widehat { { \mathcal J}}_e(x)\widehat { { \mathcal J}}_o'(x)\normordboson\right)dx,\\
\widehat{\mathcal{W}}_2=-\frac{\sqrt{2}}{3\sqrt{3}\pi i}\oint v_e(\sqrt{x})\left(\normordboson \widehat { { \mathcal J}}_e(x)\left( \widehat{{ \mathcal J}}_o'(x)^2+\frac{1}{8x^2}\right)-\frac{1}{3}\widehat { { \mathcal J}}_e(x)^3\normordboson\right)\frac{dx}{\sqrt{x}}.
\ee
If we introduce the topological expansion of the tau-function
\be
\tau_{N}({\bf t},\hbar)=1+\sum_{k=1}^\infty \hbar^k \tau_N^{(k)}({\bf t}),
\ee
then the coefficients $\tau_N^{(k)}({\bf t})$, which are polynomials in $t_k$'s, are given by a recursion
\begin{equation}
\tau_N^{(k)}=\frac{1}{k}\left(\widehat{\mathcal{W}}_{1}\,\tau_N^{(k-1)}+\widehat{\mathcal{W}}_{2}\,\tau_N^{(k-2)}\right)
\end{equation}
with the initial conditions $\tau_n^{(0)}=1$,  $\tau_n^{(-1)}=0$. This recursion is quite convenient from the computational point of view and gives a generalisation of the cut-and-join description of the Kontsevich--Witten tau-function, obtained in \cite{KontsCaJ}.

\section*{Acknowledgments}
The author is grateful to A. Mironov for useful discussions. This work was supported in part the Natural Sciences and Engineering Research Council of Canada (NSERC), by the Fonds de recherche du Qu\'ebec Nature et technologies (FRQNT) and by RFBR grants 14-01-00547 and 15-52-50041YaF.

\end{document}